# A first look at the performances of a Bayesian chart to monitor the ratio of two Weibull percentiles

Pasquale Erto

University of Naples Federico II, Naples, Italy

e-mail: ertopa@unina.it

**Abstract.** The aim of the present work is to investigate the performances of a specific Bayesian control chart used to compare two processes. The chart monitors the ratio of the percentiles of a key characteristic associated with the processes. The variability of such a characteristic is modeled via the Weibull distribution and a practical Bayesian approach to deal with Weibull data is adopted.

The percentiles of the two monitored processes are assumed to be independent random variables. The Weibull distributions of the key characteristic of both processes are assumed to have the same and stable shape parameter. This is usually experienced in practice because the Weibull shape parameter is related to the main involved factor of variability. However, if a change of the shape parameters of the processes is suspected, the involved distributions can be used to monitor their stability.

We first tested the effects of the number of the training data on the responsiveness of the chart. Then we tested the robustness of the chart in spite of very poor prior information. To this end, the *prior* values were changed to reflect a 50% shift in both directions from the original values of the shape parameter and the percentiles of the two monitored processes. Finally, various combinations of shifts were considered for the *sampling* distributions after the Phase I, with the purpose of estimating the diagnostic ability of the charts to signal an out-of-control state. The traditional approach based on the *Average Run Length*, empirically computed via a Monte Carlo simulation, was adopted.

## 1. Introduction

The need to compare two processes arises even independently from their being in statistical control or not. For example, sometime it is needed to compare the products from two different production lines to know if the two processes are making the same product simultaneously.

When two whichever processes must be compared continuously, the solution can be a monitoring chart of the ratio of specific key characteristics associated with the processes. Usually, the variability of such characteristics is modeled via skewed distribution as the Weibull (Meeker and Hamada, 1995).

Comparing the processes via the mean and variance of their key characteristics is often less meaningful than monitoring a more *operative* parameter like a specific percentile. Moreover, a small variation in mean and/or variance can hide a significant shift in the percentiles (see Padgett and Spurrier, 1990).

Especially when dealing with few data, non-Normality and available prior information, the Bayesian control charts can be an interesting alternative to traditional control charts. However, their performance investigation can result a very challenging task.

The Bayesian control charts studied in the present paper utilize a widely verified practical approach to deal with Weibull data (Erto 2005; Erto and Pallotta 2007; Hsu et al. 2011; Erto et al. 2014). We assumed that the percentiles of the two monitored processes are independent random variables. Besides, we assumed that the Weibull distributions of the key characteristic of both processes have the same and stable shape parameter. This is usually experienced in practice because the Weibull shape parameter is a characteristic of the involved main cause of variability. For example, it is related to the dispersion of flaws in the processed raw material (e.g., see Padgett et al. 1995). Thus, the shape parameter can be considered constant, even if unknown as, for example, in Nelson (1979). However, if a change to the shape parameters is suspected, their involved distributions can be used to monitor their stability.

The remainder of the paper is organized as follows. The next section briefly introduces the examined control chart. The following sections describe the procedures to test the chart performances depending on the number of used training data, poor prior information and out-of-control (OoC) data.

## 2. The control chart of the ratio of two Weibull percentiles

Given a Weibull random variable $x$, the corresponding cumulative density function is:

$$F(x;\delta,\beta) = 1 - \exp\left[-(x/\delta)^\beta\right] \qquad x \geq 0; \quad \delta, \beta > 0 \qquad (1)$$

where $\delta$ is the scale parameter and $\beta$ is the shape parameter. Let $R$ denote a specified reliability level. Then, the corresponding Weibull percentile $x_R$ can be expressed as $x_R = \delta \left[\ln(1/R)\right]^{1/\beta}$. The proposed Bayesian approach (Erto 2015) requires anticipating a numerical interval $(\beta_1, \beta_2)$ for the shape parameter $\beta$ and a likely value for the percentile $x_R$, based on previous experiments and expert opinion.

Consider two Weibull processes $x$ and $y$, that we want to monitor, and the respective first $k$ random vectors $\underline{x}_1, \ldots, \underline{x}_k$ and $\underline{y}_1, \ldots, \underline{y}_k$ of $n$ data sampled from them. Assuming that the two percentiles, $x_R$ and $y_R$, are independent random variables with the same $\beta$ shape parameter, the probability density function of their ratio $u = x_R/y_R$ is:

$$\text{pdf}\{u|\underline{x}_1,\ldots,\underline{x}_k,\underline{y}_1,\ldots,\underline{y}_k,\beta\} = \beta \frac{\Gamma[2(k \cdot n + 1)]}{[\Gamma(k \cdot n + 1)]^2} u^{\beta(k \cdot n + 1) - 1} \frac{[C(k)]^{k \cdot n + 1}}{\left[1 + u^\beta C(k)\right]^{2(k \cdot n + 1)}}. \qquad (2)$$

$$C(k) = B(k)/A(k)$$

where:

$$A(k) = a^{-\beta} + \ln\left(\frac{1}{R}\right) \sum_{i=1}^{k \cdot n} x_i^\beta, \qquad B(k) = a^{-\beta_k} + \ln\left(\frac{1}{R}\right) \sum_{i=1}^{k \cdot n} y_i^{\beta_k}. \qquad (3)$$

The estimates $\hat{\beta}_{x,i}$ and $\hat{\beta}_{y,i}$ of the shape parameter $\beta$ are the following *posterior* expectations that we can easily compute numerically:

$$\hat{\beta}_{x,k} = \text{E}\{\beta|\underline{x}_1,\ldots,\underline{x}_k\} = \frac{\int_{\beta_1}^{\beta_2} \beta^{k \cdot n + 1} a^{-\beta} \prod_{i=1}^{k \cdot n} x_i^{\beta - 1} A(k)^{-(k \cdot n + 1)} d\beta}{\int_{\beta_1}^{\beta_2} \beta^{k \cdot n} a^{-\beta} \prod_{i=1}^{k \cdot n} x_i^{\beta - 1} A(k)^{-(k \cdot n + 1)} d\beta}. \qquad (4)$$

$$\hat{\beta}_{x,k} = \text{E}\{\beta|\underline{x}_1,\ldots,\underline{x}_k\} = \frac{\int_{\beta_1}^{\beta_2} \beta^{k \cdot n + 1} a^{-\beta} \prod_{i=1}^{k \cdot n} x_i^{\beta - 1} A(k)^{-(k \cdot n + 1)} d\beta}{\int_{\beta_1}^{\beta_2} \beta^{k \cdot n} a^{-\beta} \prod_{i=1}^{k \cdot n} x_i^{\beta - 1} A(k)^{-(k \cdot n + 1)} d\beta}. \qquad (5)$$

where $\beta_1 = \hat{\beta}_{k-1}/2$ and $\beta_2 = \hat{\beta}_{k-1} \times 1.5$ in order to obtain a reasonable large symmetrical interval.

The following posterior expectations are the point estimate of $x_R$ and $y_R$:

$$\hat{x}_{R,k} = E\{x_R | \underline{x}_1, \ldots, \underline{x}_k, \bar{\beta}_k\} = \frac{\Gamma(k \cdot n + 1 - \bar{\beta}_k^{-1})}{\Gamma(k \cdot n + 1)} A(k)^{\frac{1}{\bar{\beta}_k}} \quad (6)$$

$$\hat{y}_{R,k} = E\{y_R | \underline{y}_1, \ldots, \underline{y}_k, \bar{\beta}_k\} = \frac{\Gamma(k \cdot n + 1 - \bar{\beta}_k^{-1})}{\Gamma(k \cdot n + 1)} B(k)^{\frac{1}{\bar{\beta}_k}} \quad (7)$$

where $\bar{\beta}_k$ is the average (8) of all the posterior estimates of $\beta$ accumulated up to and including the two $k^{\text{th}}$ ones:

$$\bar{\beta}_k = \frac{1}{k} \sum_{i=1}^{k} \left( \hat{\beta}_{x,i} + \hat{\beta}_{y,i} \right) / 2. \quad (8)$$

From (6) and (7) we obtain all the point estimates of the ratio $u = x_R / y_R$. Using the one-to-one transformation:

$$v = u^{\bar{\beta}_k} C(k) \quad (9)$$

we obtain the probability density function of $v$:

$$\text{pdf}\{v\} = \frac{\Gamma[2(k \cdot n + 1)]}{[\Gamma(k \cdot n + 1)]^2} \frac{v^{k \cdot n}}{[1 + v]^{2(k \cdot n + 1)}} \quad (10)$$

which is the Inverted Beta. Thus, using the inverse of the transformation (9) and given a false alarm risk $\alpha$, we can easily estimate the control limits $\text{LCL} = u_{R,\alpha/2}$ and $\text{LCL} = u_{R,1-\alpha/2}$ of the $u = x_R / y_R$ ratio as simple transformations of the percentiles, $v_{\alpha/2}$ and $v_{1-\alpha/2}$ respectively, of the Inverted Beta.

## 3. The Effect of the Number of Training Samples

It is important to note that the greater the number $m$ of the training data of the Phase I is the lower the *first* priors effects are. In fact, by accumulating training data the weight of the sampling information tends to overcome the *initial* prior information about the shape parameter $\beta$ and the percentiles $x_R$ and $y_R$ of the two processes. Consequently, as the number of the training data increases, the control limits approach a stable in-control (IC) values that express only the sampling variation. Simultaneously, the greater the number of the training data is the *stronger* (in Bayes sense) the last joint posterior of the Phase I is, since it includes the whole accumulated dataset. Because this posterior is used as prior for the following sampling, if it is excessively strong, the responsiveness of the chart toward eventual incoming out-of-control (OoC) data decreases consequently.

To highlight these features, consider the application proposed in (Erto 2015) to the data given in Huang and Johnson (2006) to compare two productions of specimens of the same (Douglas) fir tree. The parameter of interest is the 0.05 percentile of the distribution of the modulus of rupture (MOR). It is generally expected that the MOR follows the Weibull distribution (Johnson et al., 2003; Verrill et al., 2012) with a high $\beta$ parameter, say $\beta = 5$, since a close to symmetrical shape is expected in this case (Huang and Johnson 2006).

From the first manufacturing process, the estimates of the percentile $\hat{x}_{0.95}$ are obtained using samples of $n = 4$ specimens, each with cross section of $2 \times 4 = 8$ square inches. The MOR of each specimen is measured in GPa (Giga-Pascals) and reported in Table 1 divided by 10. For these specimens the value $\bar{x}_{0.95} = 2.9$ (GPa×10) is anticipated for the percentile $\bar{x}_R$ and the prior shape parameter $\bar{\beta} = 5.0$ is anticipated for $\beta$, from which a reasonable large symmetrical interval $5.0 \times (1 \mp 0.5)$ (i.e.: 2.5, 7.5) can be anticipated for $\beta$. The first $m = 10$ samples are supposed IC.

Table 1. First process: $x$ MOR (GPa×10) of samples of $n = 4$ specimens with 2×4 inches cross section.

| 3.7, 3.3, 4.9, 4.3 | 4.8, 4.6, 5.6, 4.7 | 4.8, 4.0, 4.6, 4.2 | 5.2, 4.4, 5.3, 5.0 | 4.8, 3.1, 3.9, 4.3 |
|---|---|---|---|---|
| 4.5, 4.2, 3.8, 4.1 | 3.2, 3.0, 3.6, 5.6 | 3.2, 2.4, 3.6, 4.0 | 3.0, 6.4, 4.2, 2.8 | 3.9, 3.8, 3.4, 2.5 |
| 2.9, 1.7, 3.4, 2.9 | 3.3, 3.7, 4.0, 3.3 | 4.1, 3.8, 4.4, 1.9 | 3.1, 3.7, 3.9, 2.7 | 3.1, 2.7, 3.5, 2.8 |
| 3.5, 3.9, 3.2, 4.1 | 1.6, 1.9, 3.8, 2.6 | 2.2, 3.4, 1.6, 1.8 | 3.3, 2.1, 2.9, 3.0 | 2.1, 3.6, 2.4, 3.1 |
| 2.3, 2.2, 3.6, 2.9 | 2.7, 1.9, 3.1, 3.4 | 3.5, 3.6, 0.98, 2.1 | 3.1, 1.3, 2.5, 2.3 | 4.7, 1.8, 0.85, 4.1 |

Similarly, from the second manufacturing process, the estimates of the percentile $\hat{y}_{0.95}$ are obtained using samples of $n = 4$ specimens, each with cross section of $2 \times 6 = 12$ square inches. The MOR of each specimen is measured in GPa (Giga-Pascals) and reported in Table 2 divided by 10. For these specimens a different 0.05 percentile value, $\bar{y}_{0.95} = 3.8$ (GPa×10), is anticipated (depending on the different cross section) but the same prior interval for $\beta$ (i.e.: 2.5, 7.5) is anticipated. As before, the first ten samples are supposed IC.

Table 2. Second process: $y$ MOR (GPa×10) of samples of $n = 4$ specimens with 2×6 inches cross section.

| 6.6, 4.5, 5.8, 6.5 | 6.4, 7.3, 5.6, 6.8 | 5.5, 5.7, 5.4, 5.5 | 6.2, 5.3, 4.6, 6.0 | 7.6, 6.3, 5.8, 7.1 |
|---|---|---|---|---|
| 6.1, 4.7, 5.4, 4.6 | 5.4, 3.5, 4.5, 4.5 | 3.8, 5.0, 5.5, 4.9 | 6.2, 6.2, 5.8, 5.3 | 4.7, 5.7, 4.6, 5.4 |
| 5.0, 5.4, 5.5, 5.4 | 5.1, 4.5, 3.8, 5.4 | 4.2, 3.7, 5.4, 3.6 | 5.7, 3.2, 5.1, 4.5 | 2.7, 4.7, 5.4, 6.5 |
| 4.5, 3.6, 6.0, 5.0 | 4.9, 4.7, 5.4, 4.5 | 4.6, 2.7, 4.7, 5.1 | 3.8, 5.0, 5.4, 3.9 | 4.9, 6.2, 5.0, 3.6 |
| 4.3, 4.8, 7.0, 3.8 | 4.0, 3.2, 3.9, 5.5 | 4.1, 4.2, 4.8, 3.5 | 4.2, 3.2, 2.5, 3.7 | 3.4, 3.7, 2.9, 5.1 |

The ratios $\hat{x}_{0.95}/\hat{y}_{0.95}$ of the percentiles estimated for the two processes (6) (7) give the points of the ratio charts shown in Figure 1. These charts are based on the hypothesis that $x_R$ and $y_R$ are independent random variables with the same $\beta$, as it is usually found. The anticipated value for $\beta$ is 5, as before, and the anticipated value of the ratio $u = x_{0.95}/y_{0.95}$ is assumed to be 0.76 from the ratio of the previous anticipated mean values $\bar{x}_{0.95}/\bar{y}_{0.95} = 2.9/3.8 \cong 0.76$. Using the inverse of the transformation (9) and given always the same false alarm risk, $\alpha = 0.27\%$, we obtain the control limits $LCL = u_{R,\alpha/2}$ and $UCL = u_{R,1-\alpha/2}$ as simple transformations of the percentiles, $v_{\alpha/2}$ and $v_{1-\alpha/2}$ respectively, of the Inverted Beta.

In (Erto 2015) has shown that the ratio chart does not signal any OoC state and it was concluded that the ratio $x_R/y_R$ of the percentiles of these two processes was in statistical control, that is the hypothesis of *homogeneity* of the two productions can be supported. Therefore, starting from the $(m+1)$th sample, we now produced a shift in the $y_R$ Weibull percentile by multiplying by 1.15 the data of the last 15 samples, that is all the data of the Phase II reported in Table 2. After twelve samples of the Phase II, the first chart shown in Figure 1 a) signals an OoC that indicates a lack of homogeneity of the two processes. Thus, a warning is raised and investigations can start to identify the assignable factors that cause the observed *non-proportional* effects on the two processes.

In order to highlight the considerations done at the beginning of this paragraph, we first extended the Phase I to further ten samples (of size $n = 4$) by resampling from the original set of the IC data of the first ten samples of the Table 1 and Table 2. The resulting chart is reported in Figure 1 b).

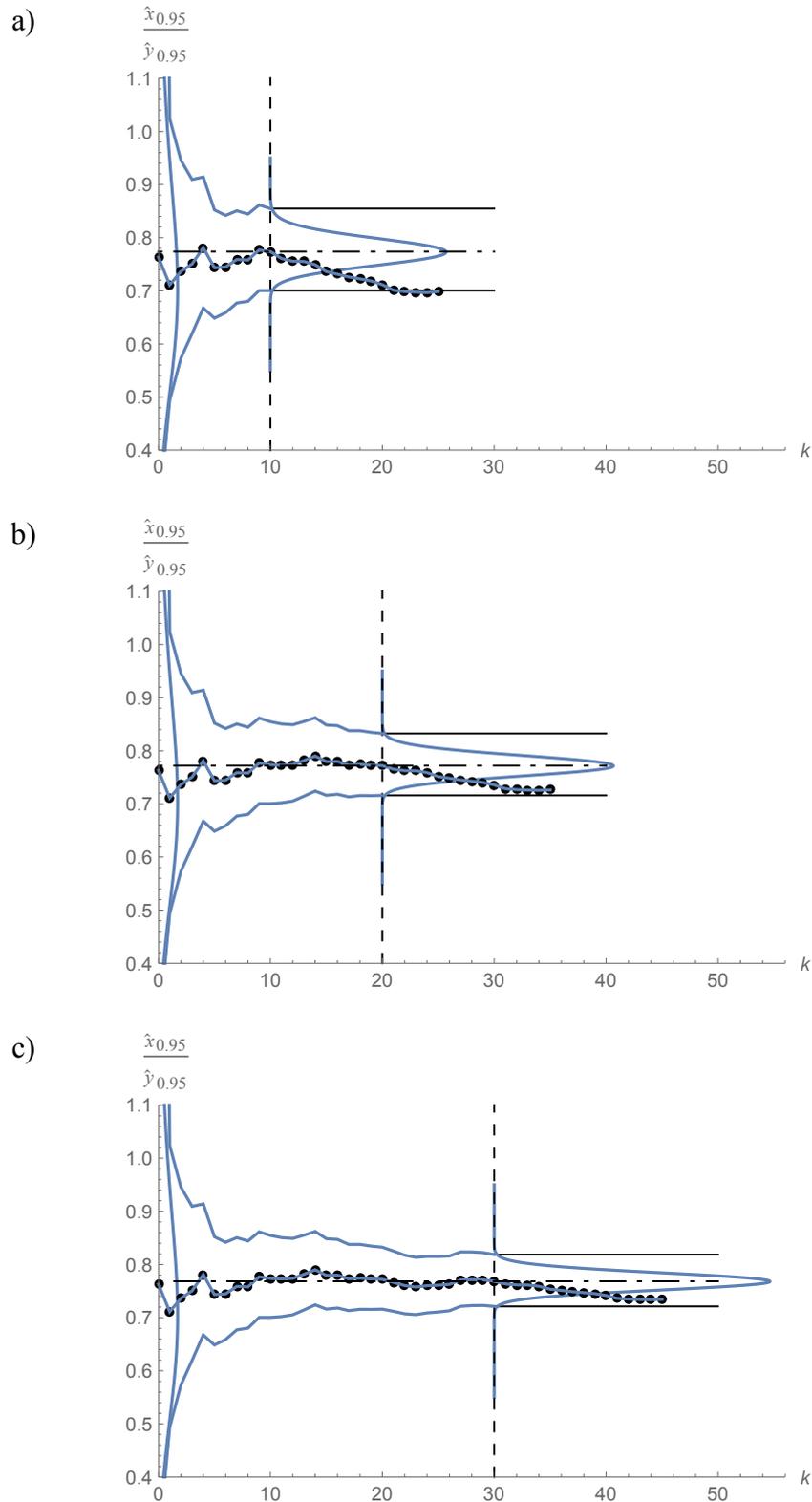

Figure 1. a) $m = 10$ implies $RL = 12$ and $UCL - LCL \cong 0.15$. b) $m = 20$ implies $RL > 15$ and $UCL - LCL \cong 0.12$  c) $m = 30$ implies $RL > 15$ and $UCL - LCL \cong 0.10$. Diagrams of the $\text{pdf}\{u|\underline{x}_k, \underline{y}_k, \beta\}$ at beginning and end of the Phase I of each chart.

The attained control interval ($UCL-LCL \cong 0.12$) is narrower than before ($UCL-LCL \cong 0.15$) but the chart does not detect the OoC (until the last available 15th sample of the Phase II). This is the effect of the stronger last joint posterior of the Phase I. More noticeable effects (Figure 1 c) are obtained by extending the Phase I to further ten samples, all resampled from the original set of the IC data as before.

If we include only the last ten IC samples in the last joint posterior of the Phase I (which works as the first prior for the following sampling of the Phase II) the responsiveness of the chart increases noticeably, as it is shown in Figure 2. In fact, the chart provides a prompt response at the $10^{th}$ and $5^{th}$ sample (after the simulated shift) when $m = 20$ and $m = 30$ respectively. Obviously, the control intervals of both b') and c') charts of Figure 2 are the same of the corresponding b) and c) charts of Figure 1.

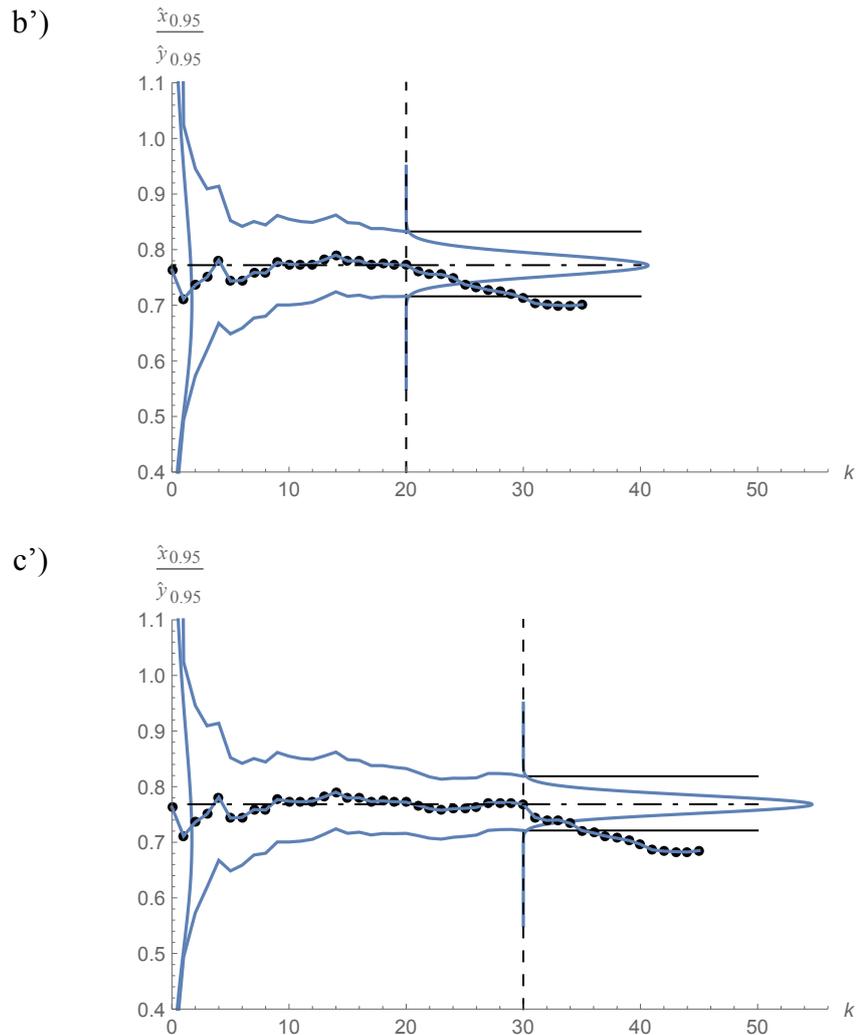

Figure 2. More responsive charts as consequence of the weaker first prior of the Phase II: b') $m = 20$ implies $RL = 10$ and c') $m = 30$ implies $RL = 5$ instead of $RL > 15$ in both cases.

In conclusion, even if exploiting as many training data as possible allows to set up the narrowest possible control interval, this advantage decreases rapidly. Besides, including too much training data (in the last joint posterior of the Phase I) reduces the detection properties of the chart during the Phase II.

## 4. The Effect of poor prior information

By considering the same applicative example of the previous paragraph, we found that even when poor prior information for $x_R$, $y_R$ and $\beta$ are adopted, the $u = x_R/y_R$ chart performances are not significantly affected. For example, we considered values of the *initial* prior information ($\bar{x}_R$, $\bar{y}_R$ and $\bar{\beta}$) changed to reflect even a 50% shift in both directions from the original unbiased values. Obviously, this does not merely imply a 50% decrease (increase) in the Weibull process mean $\mu$ and/or variance $\sigma^2$, holding the relationships:

$$\delta = x_R\left[\ln(1/R)\right]^{-1/\beta}; \quad \mu = \delta\ \Gamma(1+1/\beta); \quad \sigma^2 = \delta^2\left[\Gamma(1+2/\beta)+\Gamma^2(1+1/\beta)\right]. \quad (11)$$

We extended the Phase II to further samples (of size $n = 4$) by resampling from the original set of the Phase II samples of the Table 1 and Table 2.

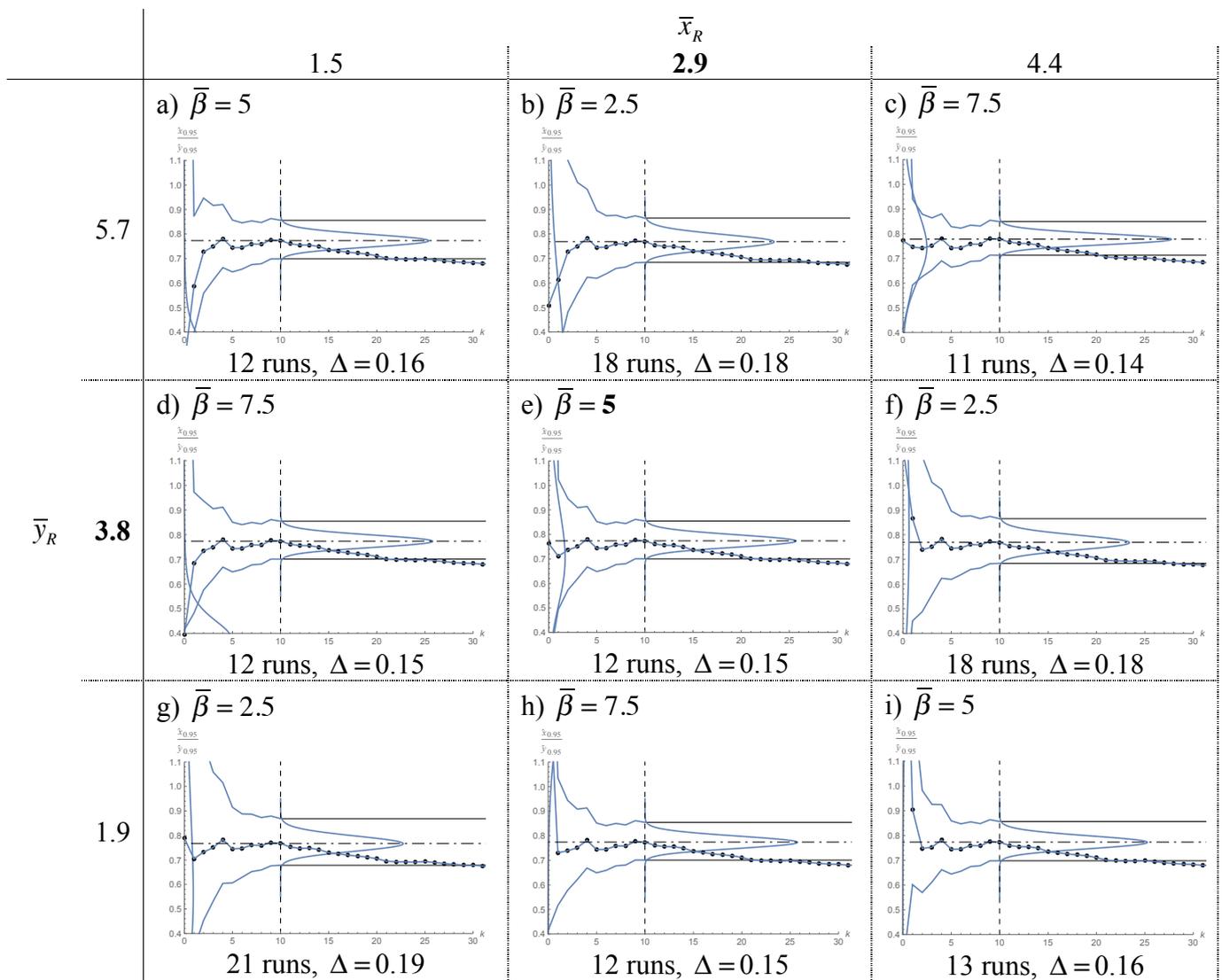

Figure 3. The effect of nine different priors on the $\hat{x}_{0.95}/\hat{y}_{0.95}$ chart performances (in boldface the unbiased prior values).

We see that even when the hypothesized prior information is very poor if compared to the baseline (see panel e in Figure 3), the Bayesian chart still shows a robust diagnostic property, in terms of OoC signals (see panels a, c, g and i of Figure 3). In fact the number of samples until the control chart signals ranges from 11 to 21. Moreover, the extent of the control limits derived from the Phase I is always close to that obtained in the baseline case, regardless of the selected prior values.

## 5. The out-of-control (OoC) performances

The OoC performances of the proposed control chart were also investigated via a Monte Carlo study. The methodology is the same used in Padgett and Spurrier (1990). We simulated $N = 1000$ cases of the proposed chart for each different OoC scenarios. For each of these, we performed a Phase I analysis by generating $m = 20$ samples of size $n = 5$ from the IC Weibull distribution and, using them, we computed the control limits and estimated the first joint prior of the Phase II. Then, we gave the Weibull distribution a percentile shift of a given magnitude, continuing to generate samples until the first point plots outside the control limits. By registering the number of samples until the chart signals, we estimated the Average Run Length (ARL) and the Standard Deviation of the Run Lengths (SDRL).

To generate the IC Weibull data, we adopted the parameter values $x_R = 1$, $y_R = 1$, $\beta = 3$ and for the prior information the unbiased prior values $\bar{x}_R = 1$, $\bar{y}_R = 1$ and $\bar{\beta} = 3$, being $R = 0.95$. The prior parameter $\beta_1$ and $\beta_2$ were obtained by multiplying $\bar{\beta}$ by 0.5 and 1.5 respectively.

To generate the OoC Weibull data, we changed one or both values $x_R = 1$ and $y_R = 1$ to $x_R^{out}$ and $y_R^o$ as indicated in the Table 3.

We tested the proposed chart in all the fourteen scenarios indicated in the Table 3, where the corresponding estimated ARL and (in brackets) SDRL are reported too. The chart shows to be able to detect even moderate shifts with satisfactory ARLs. The worst performances are obtained for the ratio $x_R^{out}/y_R^{out}$ equal to $0.5/0.8$, $0.8/1.2$, $0.5/1.0$, and for their inverses.

Table 3. Estimated Average Run Length (ARL) and, in brackets, the Standard Deviation of the Run Lengths (SDRL) of the $x_R/y_R$ control charts with in-control $ARL \cong 370$, $n = 5$, $m = 20$, $R = 0.95$.

| $x_R^{out}/y_R^{out}$ | 0.5/0.8 | 0.5/1.0 | 0.5/1.2 | 0.5/1.5 | 0.8/0.5 | 0.8/1.2 | 0.8/1.5 |
|---|---|---|---|---|---|---|---|
|  | 0.63 | 0.50 | 0.42 | 0.33 | 1.60 | 0.67 | 0.53 |
| ARL (SDRL) | 29.9 (6.4) | 13.2 (2.6) | 7.7 (1.7) | 4.3 (1.1) | 29.9 (6.4) | 13.8 (5.4) | 5.5 (1.7) |
| $x_R^{out}/y_R^{out}$ | 1.0/0.5 | 1.0/1.5 | 1.2/0.5 | 1.2/0.8 | 1.5/0.5 | 1.5/0.8 | 1.5/1.0 |
|  | 2.00 | 0.67 | 2.40 | 1.50 | 3.00 | 1.88 | 1.50 |
| ARL (SDRL) | 13.5 (2.6) | 8.9 (4.8) | 7.8 (1.6) | 13.7 (5.3) | 4.4 (1.1) | 5.6 (1.8) | 8.8 (4.7) |

## 6. Conclusion

Even in the case of poor priors and small sample sizes, the control chart still has good detection properties and enables prompt decision-making. Since the analyzed chart accumulates information, it has been shown that it can still work even starting from a limited number of small samples and/or facing moderate shifts in one or both processes.

Exploiting as many training data as possible allows setting up the narrowest possible control interval. However, this advantage decreases rapidly. Moreover, including too much training data (in the last joint posterior of the Phase I) compromises the detection properties of the chart at the beginning of its use in the Phase II. Therefore, exploiting about $80 \div 100$ training data, to set up both the control interval and the last joint posterior of the Phase I, could be the best compromise.

We see that even when the prior information is very poor the chart still shows a robust diagnostic property, in terms of number of samples until it signals. Moreover, the extent of the control limits derived from the Phase I is always close to that obtained with unbiased priors, regardless of the adopted poor prior values.

Monte Carlo tests conducted in twenty different scenarios have shown that the chart is able to detect even moderate shifts with acceptable ARLs. The worst performances are obtained when only one process is given a moderate percentile shift, leaving the other unchanged.